\documentclass{ws-p8-50x6-00}

\begin{document}

\title{Microscopic and Bulk Spectra of Dirac Operators from 
Finite-Volume Partition Functions}

\author{G. Akemann}

\address{Max-Planck-Institut f\"ur Kernphysik\\Saupfercheckweg 1, 
D-69117 Heidelberg, Germany}

\maketitle

\abstracts{The microscopic spectrum of the QCD 
Dirac operator is shown to obey random matrix model statistics in the bulk 
region of the spectrum close to the origin using finite-volume partition 
functions.}

\section{Introduction}

The spectral density $\rho(\lambda)$ 
of the Dirac operator in QCD contains interesting informations as it  
is for example directly proportional to the chiral condensate $\Sigma$
at the origin through the Banks-Casher relation~\cite{BC}. 
While the full spectrum is only accessible numerically in QCD on the lattice
many analytic results for 
parts of the spectrum have been obtained during the past years using random 
matrix theory (RMT), chiral perturbation theory and finite-volume partition 
functions (for a recent review see~\cite{Jac}). Within these results two
different regimes have been investigated: 
(i) {\it unscaled} macroscopic correlations 
and (ii) microscopic correlations between eigenvalues {\it rescaled} by the 
mean spectral density. In the region (ii) one furthermore has to distinguish
between scaling at the {\it origin} and in the {\it bulk} of the spectrum. 
In the macroscopic regime (i) the slope of the spectral density at the origin
\begin{equation}
\rho^{\prime}(\lambda=0)\ =\ \frac{\Sigma^2}{16\pi^2F_\pi^4}
\frac{(N_f-2)(N_f+\beta)}{\beta N_f} ,
\end{equation} 
has been calculated by Smilga and Stern~\cite{SSt} for 
fundamental $SU(N_c\geq3)$ fermions $(\beta=2)$ and very recently by 
Toublan and Verbaarschot~\cite{DV99}  for fundamental $SU(2)$ 
fermions $(\beta=1)$ and adjoint fermions $(\beta=4)$ with any gauge group. 
Here $F_\pi$ denotes the pion decay constant. 

\noindent

\noindent
Coming to the rescaled correlations (ii) at the origin all higher 
order correlation functions of scaled eigenvalues $x=\lambda/\Sigma V$ are 
known. They are given analytically as functions of the chiral condensate  
$\Sigma$ and scaled sea-quark masses $\mu_f=m_f/\Sigma V$ (for details 
see~\cite{Jac}). This description holds in the limit of Leutwyler and 
Smilga~\cite{LS} $1/\Lambda_{QCD}\ll V^{1/4}\ll 1/m_\pi$, where $m_\pi$ is the 
pion mass, up to a certain scale $\lambda_{Th}\sim V^{1/2}$ called Thouless 
energy~\cite{TH} 
in analogy to disordered systems. Within this region the analytic description 
provides an exact nonperturbative limit of QCD, dealing,  
however, with an unphysically heavy pion that does not fit 
into the finite-volume of the system.

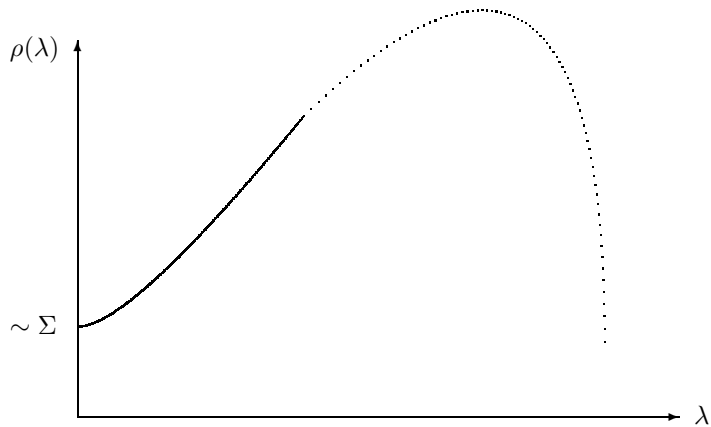
\begin{figure}[t]
%\figurebox{20pc}{15pc}{} % to have a box alone
\unitlength1cm
\begin{picture}(12.2,5.5)
\put(1.0,0.3){\vector(1,0){8}}
\put(9.2,0.2){$\lambda$}
\put(1.0,0.3){\vector(0,1){5}}
\put(0.1,5.1){$\rho(\lambda)$}
\put(0.1,1.4){$\sim\Sigma$}
\qbezier(1.0,1.5)(1.7,1.5)(4.0,4.3)
\qbezier[80](4.0,4.3)(8.0,8.2)(8.0,1.3)
\end{picture} 
%\epsfxsize=10pc 
% will enlarge or reduce the postscript figures based on the xsize
%\epsfbox{xxx.eps} % postscript image file name
\caption{The unscaled spectral density plotted schematically
(compare Ref.$^8$).\label{fig:rho}}
\end{figure}

\noindent 
Away from the origin much less is known about correlations on the 
microscopic scale. Up to now it has been found only empirically that also 
there correlations follow a (non-chiral) RMT
description~\cite{HKV,GMMW}. The aim of the results presented here is to 
understand analytically why this is possible. Therefore, the correlation 
functions are expressed in terms of finite-volume partition 
functions~\cite{AD} and explored in the bulk region but still in the vicinity 
of the origin. The region of validity is therefore given by the one of 
finite-volume partition functions. It does not hold all the way up to the 
unphysical turn-over in the spectrum (dotted line in Fig.~\ref{fig:rho})
which is due to cut-off and finite-size 
effects on the lattice. This intermediate region remains unexplained so far.

\section{Correlation functions from finite-volume partition functions}
Before taking the bulk limit we have to recall how the correlators can be 
expressed using the Leutwyler and Smilga partition functions in the 
microscopic limit at the origin. The 
generation function for the correlators is the kernel of orthogonal 
polynomials. For gauge group $SU(N_c\geq 3)$ with fermions in the fundamental 
representation it is given by~\cite{AD}
\bea
K_S(x,y;\{\mu_{{f}}\}) &=& c\sqrt{xy}\prod_{f=1}^{N_{f}}
\sqrt{(x^2+\mu_f^2)(y^2+\mu_f^2)}~\frac{
{\cal Z}_{\nu}^{(N_{f}+2)}(\{\mu_{{f}}\},ix,iy)}{
{\cal Z}_{\nu}^{(N_{f})}(\{\mu_{{f}}\})} 
\label{eq:ker} \\
{\cal Z}_{\nu}^{(N_{f})}(\{\mu_{{f}}\}) &=& 
\frac{\det {\cal A}(\{\mu_f\})}{\Delta(\mu_f^2)} \ , 
\ \ {\cal A}_{ij} ~=~ \mu_i^{j-1}I_{\nu+j-1}(\mu_i) \ ,
\label{eq:Z}
\eea
where ${\cal Z}_{\nu}^{(N_{f})}$ is the partition function 
of $N_f$ flavors with rescaled masses $\mu_f$, $\nu$ zeromodes
and $\Delta$ being the Vandermonde determinant.
The rescaled correlation functions can then be obtained as 
\begin{equation}
\rho_S^{(N_{f},\nu)}(x_1,\ldots,x_k;\{\mu_{{f}}\}) \ =\
\det_{1\leq a,b\leq k} 
K_S(x_a,x_b;\{\mu_{{f}}\}) \ .
\label{eq:MM}
\end{equation}
This description is equivalent to a unitary chiral RMT
description $(\beta=2)$.
Since we want to go to the bulk of the spectrum we should expect that the 
chiral properties no longer play a role for the correlations. We will find 
that this is indeed the case even in the vicinity of the origin. It is 
therefore instructive to compare to 3-dimensional QCD with flavor
symmetry breaking, which is described by non-chiral RMT~\cite{VZ}.
The corresponding kernel can be written in the same way as in 
Eq.~(\ref{eq:ker})~\cite{AD} after dropping the prefactor $c\sqrt{xy}$, 
where the product now runs over $N_f/2$ flavors occurring in pairs
$\pm\mu_f$. The corresponding partition function reads
\begin{equation}
{\cal Z}^{(N_{f})}(\{\mu\}) =
\det\left(\begin{array}{ll}
A(\{\mu_f\}) & A(\{-\mu_f\})\\
A(\{-\mu_f\})  & A(\{\mu_f\})
\end{array}\right)
\Delta(\{\mu\})^{-1} , 
A_{ij}= (\mu_i)^{j-1} e^{\mu_i}.
\label{eq:Z3}
\end{equation}

\section{The bulk limit}
In RMT it is known for massless flavors how to ``invert'' the origin
scaling limit where $V\to\infty$, $\lambda\to0$ and $x=\lambda/\Sigma V$
is kept fixed:
\begin{equation}
\lim_{\begin{array}{ll}
x,y\to\infty\\ x-y=O(1)
\end{array}} 
K_{\mbox{\tiny origin}}(x,y) \ =\ K_{\mbox{\tiny bulk}}(x,y) \ =\
c'\ \frac{\sin(x-y)}{x-y} \ .
\label{eq:OB}
\end{equation}
It takes us from the Bessel-kernel at the origin (Eq.~(\ref{eq:ker}) for
$m_f\!=\!0$)
to the sine-kernel in the bulk. We will see in the following that the same
result holds including the masses. In RMT the translational invariance of the
spectrum then implies that the sine-kernel holds everywhere in the bulk.
Since we cannot make such a strong statement about the QCD Dirac spectrum
we will be restricted to the domain of validity of finite-volume
partition functions.

\noindent
In taking the bulk limit of the full kernel Eq.~(\ref{eq:ker}) we also have to
scale back all masses $\mu_f\to\infty$ at the same time since they would
otherwise drop out trivially. Using the asymptotic expansion of Bessel
functions we obtain
\bea
\lim_{\begin{array}{ll}
x,y,\mu_f\to\infty\\ \mbox{\tiny differences}\ O(1)
\end{array}} 
K_S(x,y;\{\mu_{{f}}\}) &=& 
c'\prod_{f=1}^{N_f}\frac{(\mu_f^2+xy)}{\sqrt{(x^2+\mu_f^2)(y^2+\mu_f^2)}}
\frac{\sin(x-y)}{x-y}  .
\label{eq:kerB}
\eea
Expanding the prefactor around one of the arguments it is unity up to higher 
orders and thus we find the same result as in the massless
case Eq.~(\ref{eq:OB}). Taking the same limit of the non-chiral theory
Eqs.~(\ref{eq:ker}),(\ref{eq:Z3}) we obtain precisely the same result and thus
find, that the chiral properties are lost in the bulk even in the vicinity
of the origin. Another consequence is that in the bulk 
any mass scale given by the sea-quarks drops out on the microscopic scale.

\noindent
In Ref.~\cite{GMMW} it has been stated 
that the Thouless energy scale at the origin
is larger or equal to the one in the bulk.
Looking at the bulk but close to the origin it is clear that if we take
$\lambda_1<\lambda_{Th}$ and $\lambda_2>\lambda_{Th}$ we are outside the 
domain of validity 
although it may still be $|\lambda_1-\lambda_2|<\lambda_{Th}$.

\section*{Acknowledgments}
Helpful discussions with P.H. Damgaard, T. Guhr, A.D. Jackson, 
J.J.M. Verbaarschot and H.A. Weidenm\"uller are gratefully acknowledged. 
Furthermore I wish to
thank the organizers for the stimulating workshop.

\end{document}